\documentclass[journal]{IEEEtran}
\usepackage{bm}
\usepackage{epsf}
\usepackage{times}
\usepackage{float}
\usepackage{graphicx}
\usepackage{amsmath,amssymb,amsthm}
\usepackage{color}
\usepackage{cite}
\usepackage{subcaption}
\usepackage{algorithm}
\usepackage{algorithmic}

\usepackage{float} 
\allowdisplaybreaks
\makeatletter
\newcommand*{\rom}[1]{\expandafter\@slowromancap\romannumeral #1@}
\makeatother

\newtheorem{remark}{Remark}
\newtheorem{assumption}{Assumption}
\newtheorem{proposition}{Proposition}
\newtheorem{definition}{Definition}
\graphicspath{{subPert/}}

\def \mcalN{\mathcal{N}}

\def\bs{\bm s}
\def\bx{\bm x}
\def\bX{\bm X}
\def\by{\bm y}
\def\bW{\bm W}
\def\br{\bm r}

\def \bC{\bm C}
\def \bPC{\bm {PC}}
\def \bP{\bm P}

\def \bB{\bm B}
\def \bI{\bm I}
\def \blambda{\bm \lambda}

\DeclareMathOperator*{\tsum}{\textstyle\sum}

\begin{document}

\title{
{Privacy-Preserving Distributed Processing:\\
Metrics, Bounds, and Algorithms}
}
\author{Qiongxiu Li, Jaron Skovsted Gundersen, Richard Heusdens
and Mads~Gr\ae sb\o ll~Christensen
\thanks{Q. Li and M. G. Christensen are with the Audio Analysis Lab, CREATE, Aalborg University, Rendsburggade 14, Aalborg, Denmark (emails: \{qili,mgc\}@create.aau.dk).}
\thanks{ J. S. Gundersen is with the Department of Mathematical Sciences, Aalborg University, Skjernvej 4A, Aalborg, Denmark (e-mail:
jaron@math.aau.dk).}
\thanks{R. Heusdens is with the Netherlands Defence Academy (NLDA), Het Nieuwe Diep 8, 1781 AC Den Helder, The Netherlands, and with the Faculty of Electrical Engineering, Mathematics and Computer Science, Delft University of Technology, Mekelweg 4, 2628 CD Delft, The Netherlands (email: r.heusdens@\{mindef.nl,tudelft.nl\}).}
}
\maketitle
\begin{abstract}
Privacy-preserving distributed processing has recently attracted considerable attention. It aims to design solutions for conducting signal processing tasks over networks in a decentralized fashion without violating privacy. Many algorithms can be adopted to solve this problem such as differential privacy, secure multiparty computation, and the recently proposed distributed optimization based subspace perturbation. However, how these algorithms relate to each other is not fully explored yet. In this paper, we therefore first propose information-theoretic metrics based on mutual information. Using the proposed metrics, we are able to compare and relate a number of existing well-known algorithms. We then derive a lower bound on individual privacy that gives insights on the nature of the problem.  To validate the above claims, we investigate a concrete example and compare a number of state-of-the-art approaches in terms of different aspects such as output utility, individual privacy and algorithm robustness against the number of corrupted parties, using not only theoretical analysis but also numerical validation. Finally,  we discuss and provide principles for designing appropriate algorithms for different applications.
\end{abstract}


\section{Introduction}
Big data is accompanied by big challenges. Currently, data are collected and simultaneously stored on various local devices, such as phones, tablets and wearable devices \cite{anderson2015technology,poushter2016smartphone}. In these cases, three critical challenges exist in processing such large amounts of data: (1) the emerging demand for distributed signal processing tools, as these devices are distributed in nature and often rely on wireless communication to form a network that allows devices to cooperate for solving a problem; (2) the requirement for both
computational and communication efficient solutions, due to the fact  that these devices are usually resource-constrained, for example in wireless sensor networks; and (3) privacy concerns, as sensors from these devices, such as GPS and cameras, usually contain sensitive personal information.  Consequently, having efficient privacy-preserving distributed processing solutions, which are able to address the privacy concerns, is highly important and usually requires interdisciplinary research across fields such as distributed signal processing, information theory and cryptography.

Before describing related studies, we first introduce an important concept called security model. There are two primary types of security models: (1) computational security, in which the adversary is assumed to be computationally bounded such that it cannot decrypt a secret efficiently (i.e., in polynomial time) and (2) information-theoretic security, in which the adversary is assumed to be computationally unbounded but does not have sufficient information for inferring the secret. In this paper we focus on information-theoretic security since it assumes a stronger adversary and is more efficient in terms of both communication and computational demands \cite{lagendijk2013encrypted}. 
\subsection{Related works}
Many information-theoretic approaches have been proposed for addressing privacy issues in various distributed processing problems like distributed average consensus \cite{li2019privacyA,gupta2017privacy,gupta2019statistical,Jane2020ICASSP,kefayati2007secure,huang2012differentially,nozari2017differentially,manitara2013privacy,mo2017privacy,he2019privacy,braca2016learning,hale2015differentially,hale2018cloud}, distributed least squares \cite{tjell2020privacy,Jane2020LS}, distributed optimization \cite{tjell2019privacy,huang2015differentially,han2016differentially,nozari2018differentially,zhang2016dynamic,zhang2018recycled,zhang2018improving,xiong2020privacy,Jane2020Arxiv} and distributed graph filtering \cite{Jane2020GSP}. These approaches can be broadly classified into three classes. The first two classes combine distributed signal processing with commonly used cryptographic tools, such as secure multiparty computation (SMPC) \cite{damgaard2012multiparty,Cramer2015}, and privacy primitives, such as differential privacy (DP) \cite{dwork2006,dwork2009differential}, respectively. The third class directly explores the potential of existing distributed signal processing tools for privacy preservation, such as distributed optimization based subspace perturbation (DOSP) \cite{Jane2020ICASSP,Jane2020LS,Jane2020Arxiv}. Among these approaches, SMPC aims to securely compute a function over a number of parties' private data without revealing it. DP, on the other hand, is defined to ensure that the posterior guess relating to the private data is only slightly better (quantified by $\epsilon$) than the prior guess. While DOSP protects the private data by inserting noise in a specific subspace determined by the graph topology.  

There are three challenges in addressing the privacy issues for distributed processing. (1) There is no generic framework that is able to relate and quantify all existing algorithms, because each approach, e.g., SMPC, DP or DOSP, has its own metrics and features. Additionally, there are also many cases, for example distributed average consensus, in which SMPC, DOSP and DP are exclusive with respect to each other, e.g., a SMPC based algorithm can never be differentially private.  Therefore, it is very difficult to choose an appropriate algorithm given a specific application at hand. 
(2) Apply these approaches directly to distributed processing does not always guarantee the performance. As SMPC and DP were not originally defined in the context of distributed processing, there are cases where they cannot protect the private data from being revealed to others.
As an example,  a perfect SMPC protocol does not necessarily prevent privacy leakage and a perfect DP based approach ($\epsilon=0$) does not imply zero information leakage if the private data are correlated \cite{kifer2011no}.
(3) It is very challenging to analyze and visualize the information-theoretical results.  Due to the fact that distributed processing algorithms are usually iterative, it is thus very complex to analytically track the privacy analysis over all iterations. In addition to this, visualization of related theoretical results, which will help to validate and understand the algorithm performances, is also rarely explored in the literature. 
In this paper, we attempt to overcome the above mentioned challenges.
\subsection{Paper contributions}
The main contributions are summarized below.
\begin{itemize}
    \item To the best of our knowledge, this is the first paper proposing formal information-theoretic metrics that are able to quantify output utility, individual privacy and algorithm robustness in the context of privacy-preserving distributed processing. The proposed metrics are general, where well-known frameworks such as DP and SMPC can be considered as special cases.
    \item We derive both a lower bound on individual privacy and a condition that ensures that DP and SMPC/DOSP are mutually exclusive. These results not only help to understand the nature of the problem but also gives guidance on designing algorithms. 
    \item By applying into a concrete example, we are able to analyze, quantify, compare, and understand the nature of a number of different privacy-preserving algorithms including  DP, SMPC and DOSP. In addition, we also visualize all the information-theoretical results with numerical validations.
\end{itemize}

\subsection{Outline and notation}
This paper is organized as follows. Section \ref{sec.basic} states the problem to be solved. Section \ref{sec.smpc}  briefly reviews SMPC and DP, and then discusses their limitations in quantifying privacy-preserving distributed processing protocols. Section \ref{sec.propMetric} introduces the proposed metrics and relate them to both DP and SMPC. Additionally, a lower bound on individual privacy is given.  Sections \ref{sec.consensus1} and \ref{sec.consensus2} describe a concrete example, i.e., distributed average consensus. The former section defines the problem and shows that traditional approaches leak private information, while the latter section first presents a theoretical result for achieving privacy-preservation and then analyzes existing privacy-preserving distributed average consensus algorithms using the proposed metrics. Comparisons, numerical results, and discussions are given in Section \ref{sec.numRes}, and Section \ref{sec.conclu} concludes the paper. 

We use lowercase letters $(x)$ for scalars, lowercase boldface letters $(\bx)$ for vectors, uppercase boldface letters $(\bX)$ for matrices, overlined uppercase letters $(\bar{X})$ for subspaces, calligraphic letters $(\mathcal{X})$ for arbitrary sets and $|\cdot|$ for the cardinality of a set. Uppercase letters $(X)$ denote random variables having realizations $x$. 
$ \operatorname{span}\{\cdot\}  \text { and }\operatorname { null }\{\cdot\}
$ denote the span and nullspace of their argument, respectively.
$(\bX)^\dagger$ and $(\bX)^{\top}$ denote the Moore-Penrose pseudo inverse and transpose of $\bX$, respectively.
$x_{i}$ denotes the $i$-th entry of the vector $\bx$ and $\bm X_{ij}$ denotes the $(i,j)$-th entry of the matrix $\bX$.
$\boldsymbol{0}$, $\boldsymbol{1}$ and $ \bm I$ denote the vectors with all zeros and all ones, and the identity matrix of appropriate size, respectively.
\section{Preliminaries} \label{sec.basic} 
In this section, we introduce the problem setup and the adversary models. We then state the three main requirements to be addressed.
\subsection{Privacy-preserving distributed processing over networks} \label{subsec.netwSetup}
A network can be modelled as a graph $\mathcal{G}=\{\mcalN,\mathcal{E}\}$ where $\mcalN=\{1,\ldots,n\}$ denotes the set of $n$ nodes and $\mathcal{E} \subseteq \mcalN \times \mcalN$ denotes the set of $m$ edges. Let $\mcalN_{i}=\{j\mid {(i,j)\in \mathcal{E}\}}$ denote the neighborhood of node $i$ and $d_i=|\mcalN_{i}|$. Denote $e_l=(i,j) \in \mathcal{E}$, where $ l\in \{1,\ldots,m\}$, as the $l$-th edge, and let $\bB \in \mathbb{R}^{m \times n}$ be the graph incidence matrix defined as $\bB_{li}=1$, $\bB_{lj}=-1 $ if and only if $(i,j)\in \mathcal{E}$ and $i<j$. Assume each node $i$ has private data $s_i$ and let $\bs=[s_1,\ldots,s_n]^{\top}$. Note that for simplicity, $s_i$ is assumed to be scalar but the results can easily be generalized to arbitrary dimensions.  

The goal of privacy-preserving distributed processing over a network is to compute a function
\begin{align}
   f: \mathbb{R}^{n}\mapsto \mathbb{R}^{n},  \by = f(\bs),
\end{align}
in a distributed manner without revealing each node's private data $s_{i}$ to other nodes, where $y_i$ denotes the desired output of node $i$. By a distributed manner we mean that only data exchange between neighboring nodes is allowed. 
\subsection{Adversary model}
An adversary model is used to evaluate the robustness of the system under different security attacks. The adversary works by colluding a number of nodes, and it aims to conduct certain malicious activities such as inferring the private data. These colluding nodes are referred to as corrupted nodes, and the others are called honest nodes. 
In this paper, we consider two types of adversary models: passive and eavesdropping. The passive adversary model is a typical model to be addressed in distributed networks \cite{bogdanov2008sharemind}. It assumes 
that the corrupted nodes are honest-but-curious, that is, these corrupted nodes follow the algorithm instructions but will share information together to infer the private data of the honest nodes.

An eavesdropping adversary, on the other hand, is assumed to listen on all communication channels between nodes with the purpose of inferring the private data of the honest nodes.
The eavesdropping adversary model is relatively unexplored in the context of privacy-preserving distributed processing. Indeed, many SMPC based approaches, such as those based on secret sharing \cite{li2019privacyS,tjell2019privacy,tjell2020privacy}, assume that all messages are transmitted through securely encrypted channels \cite{dolev1993perfectly} such that the communication channels cannot be eavesdropped. However, channel encryption is computationally demanding and is, therefore, very expensive for iterative algorithms, such as those considered here. 

\subsection{Main requirements}
We identify three important factors to be considered when designing a privacy-preserving distributed processing algorithm:
\begin{enumerate}
    \item Output utility: at the end of the algorithm, each node would like to obtain its desired output $y_i$. As the goal is to not compromise the accuracy of output by considering privacy, we thus consider the output of traditional distributed processing approaches, i.e.,  without any privacy concern, as the baseline.  Hence, the desired output is defined as the one computed by a traditional non-privacy-preserving algorithm.
    \item Individual privacy: during the entire algorithm execution, each node wants to prevent its private data $s_i$ from being revealed to others. 
    \item Algorithm robustness:
    the algorithm should be able to preserve privacy in the presence of a large number of corrupted nodes.
\end{enumerate}
\section{Secure multiparty computation and differential privacy} \label{sec.smpc}
This section briefly introduces two widely used techniques for privacy-preservation: SMPC and DP, and discuss their limitations in quantifying the performance of privacy-preserving distributed processing algorithms. 
\subsection{Secure multiparty computation} \label{subsec.mecSMPC}
An important concept in SMPC is the definition of an ideal world, in which a trusted third party (TTP) is assumed to be available.  A TTP works by first computing the function result $\by = f(\bs)$ after
collecting all private data from each node and then sending the desired outputs $y_i$ to each and every node. This scenario is considered secure since a TTP is assumed non-corrupted. However, there is a distinction between security and privacy.  We remark that an ideal world does not necessarily guarantee zero privacy leakage. This is because the passive adversary always has the knowledge of the private data and desired output of the corrupted nodes. This knowledge can leak information about the private data $s_i$ of honest node $i$, which can be quantified by 
\begin{align}\label{eq.smpcImp}
    I(S_i;\{S_j,Y_j\}_{j\in \mcalN_c}),
\end{align}
where $\mcalN_c$ denotes the set of corrupted nodes and  $I(\cdot\,;\,\cdot)$ denotes mutual information \cite{cover2012elements}. Apparently, this information loss is not necessarily zero and it is indeed dependent on several factors such as the function type and whether the private data are correlated or not. 

The motivation of SMPC comes from the fact that in practice a third party might not be available or trustworthy. The goal of SMPC is thus to design a protocol that can replace a TTP.  Therefore,
a SMPC protocol is considered to be perfect whenever the adversary does not learn more about each honest node's private data than what is already revealed in an ideal world as quantified in \eqref{eq.smpcImp}. 
Again, a perfect SMPC protocol does not imply zero information leakage (but only means that it successfully replaces a TTP).

As an example in which SMPC violates individual privacy, consider the situation in which $\by$ is a permuted version of the private data $\bs$. That is, $y_{i}=s_{i\,-\,1\!\!\!\mod\! n}$. Then, if node $i$ is corrupted, the private data of node $i\,-\,1\!\!\!\mod\! n$ will be revealed regardless of the SMPC protocol.
Therefore, we conclude that SMPC based approaches might not preserve privacy at all, and using SMPC metrics are thus insufficient for quantifying the performance of privacy-preserving distributed processing algorithms.
  
\subsection{Differential privacy} \label{subsec.dp}
DP is a protocol that can be used when recruiting a person to participate in distributed processing. DP aims to protect this person's privacy in an extreme case where all other persons are assumed to be not trust-worthy. That is,  there are $n-1$ corrupted nodes and only one honest node, say node $i$, i.e., $\mcalN_c=\mcalN \setminus \{i\}$. To emulate such a scenario, let $\bs'\in \mathbb{R}^{n}$ be an adjacent vector of $\bs$ where $\forall j\in \mcalN_c: s'_j=s_j$ and $s'_i\neq s_i$. Let $\mathcal{Y}$ denote the output range of a function $f$. Given $\epsilon \geq 0$, an algorithm achieves $\epsilon$-DP if for any pair of adjacent vectors $\bs$ and $\bs'$, and for all sets $\mathcal{Y}_s\subseteq \mathcal{Y}$, we have 
\begin{align}\label{eq.oriDP}
 \ P(f(\bs)\in \mathcal{Y}_s )\leq e^{\epsilon}P(f(\bs')\in \mathcal{Y}_s).
\end{align}
Note that the adversary model considered in DP is different from the previously defined passive adversary model: the former assumes that the adversary knows the function outputs given two adjacent vectors $\bs$ and $\bs'$. While the latter assumes that both the private data and function outputs of the corrupted nodes are known to the adversary.  Consequently,  $\epsilon=0$ does not imply zero information leakage if the private data are correlated since $I(S_i;\{Y_j,S_j\}_{j\in \mcalN_c})\neq 0$.  We conclude that applying DP directly to distributed processing does not always guarantee privacy and DP is not sufficient for quantifying the privacy.

\section{Proposed metrics and bounds} \label{sec.propMetric}
This section starts by introducing the proposed metrics and in particular we present a lower bound on individual privacy. After that, we will relate the proposed metric to both SMPC and DP. 
The proposed metrics $(u_i, \rho_i,k_i)$ are for each individual node and are defined as follows. 
\subsection{Output utility $u_i$}
Information-theoretic approaches achieve privacy-preservation mainly by using data obfuscation/perturbation through noise insertion. Let $\br \in \mathbb{R}^{n}$ denote the inserted noise realization.
The estimated function output is given by
\begin{align*}
    \hat{\by}=f(\bs,\br).
\end{align*}
Note that the desired function output $\by$ can be seen as a special case where no noise is inserted: 
\begin{align*}
    \by=f(\bs,\bm 0).
\end{align*}
To define an information-theoretic utility measure, mutual information has been widely adopted in the literature \cite{duchi2013local,kairouz2014extremal}. Here, we also use mutual information to define the output utility:
\begin{align}
    \forall i\in \mathcal{N}: u_i=I(Y_i;\hat{Y}_i).
\end{align}
We can see that $u_i \in [0,I(Y_i;Y_i)]$ and $u_i=I(Y_i;Y_i)$ means full utility. In theory, if $Y_i$ is a discrete random variable we have $I(Y_i;Y_i)=H(Y_i)$ where $H(\cdot)$ denotes the Shannon entropy and $I(Y_i;Y_i)= +\infty$ if $Y_i$ is a continuous random variable. Note that in practice we only deal with the former case as computers can only process discretized data.
\subsection{Individual privacy $\rho_i$}
When defining privacy, the $\epsilon$-DP shown in \eqref{eq.oriDP} has been widely used because it is a worst-case metric that provides a strong privacy assurance in any situation, e.g., for all prior distributions of the private data. However, besides the problem of not working for correlated data, such strong assurances can be very difficult to guarantee in practice \cite{gotz2011publishing,haeberlen2011differential,korolova2009releasing}. In addition, this worst-case privacy leakage can in practice be quite far from the typical leakage of the average user \cite{lopuhaa2019information}. For these reasons, many relaxations of $\epsilon$-DP have been proposed \cite{cuff2016differential,dwork2016concentrated,dwork2006our,mironov2017renyi}. In this paper we will deploy mutual information as the individual privacy metric; it is a relaxation of DP since it is an average metric. To quantify the individual privacy of honest node $i \in \mathcal{N}_{h}$ where $\mathcal{N}_{h}=\mathcal{N} \setminus \mathcal{N}_{c}$ denote the set of honest nodes, we first denote $\mathcal{V}_i$ as the set of random variables which contains all the information collected by the adversary for inferring the private data $s_i$. The individual privacy is thus defined as
\begin{align}
   \forall i\in \mathcal{N}_h: \rho_i=I(S_i,\mathcal{V}_i).
\end{align}
\subsubsection{lower bound on individual privacy}
The individual privacy $\rho_i$ is lower bounded by 
\begin{align}\label{eq.lbPro}
 \rho_{i,\min}=I(S_i;\{S_j,\hat{Y}_j\}_{j \in \mcalN_{c}}).
\end{align}
This is due to $\{S_j,\hat{Y}_j\}_{j \in \mcalN_{c}}$ being the minimum knowledge available to the adversary. Let $\mathcal{V}_{i,\min}=\{S_j,\hat{Y}_j\}_{j \in \mcalN_{c}}$.  Imagine the whole distributed processing as a blackbox with $\bs$ as inputs and $\hat{\by}$ as the output. Recall the definition of the passive adversary model, by controlling a number of corrupted nodes the adversary always has the knowledge of their private data and estimated outputs, regardless of the algorithm adopted (i.e., independent of the information flow within the blackbox). Therefore, we conclude that $\mathcal{V}_{i,\min} \subseteq \mathcal{V}_i$, thus also $\rho_{i,\min} \leq \rho_i$. Notably, this lower bound becomes \eqref{eq.smpcImp} in SMPC when full utility  $\hat{\by}=\by$ is achieved.

Here we propose a new definition of perfect individual privacy in the context of distributed processing. Intuitively, perfect individual privacy means zero information leakage, i.e., $\rho_i=0$. However, due to the fact that the lower bound $\rho_{i,\min}$ is not necessarily zero since it is dependent on several factors such as the type of function, the estimated output and the number of corrupted nodes,  it is thus impossible to achieve zero information loss if $\rho_{i,\min} > 0$. Therefore, we introduce the following definition. 
\begin{definition}\label{def.1} (Perfect individual  privacy in the context of privacy-preserving distributed processing.)
Given $\rho_{i,\min}\in [0,I(S_i;S_i))$, a privacy-preserving algorithm is perfect (i.e., achieves perfect individual privacy) if it reaches the lower bound, i.e., $\rho_i=\rho_{i,\min}$.
\end{definition}

\subsection{Algorithm robustness $k_i$} The algorithm robustness is quantified by $k_i \in \{0,\ldots, n-1\}$, which measures the maximum number of corrupted nodes that can be tolerated under a passive adversary. $k_i = n-1$ means the algorithm is able to protect the private data $s_i$ from being revealed even if all other nodes in the network are corrupted. Note that the algorithm robustness is defined under a passive adversary model. For the case of an eavesdropping adversary, we will address it by the cost of channel encryption.

\subsection{Linking the proposed metrics to SMPC and DP}
The proposed metrics $(u_i,\rho_i,k_i)$ are closely related to the well-known SMPC and DP:
\begin{enumerate}
    \item They reduce to perfect SMPC when $\forall i\in \mathcal{N}: u_i=I(Y_i;Y_i)$ and $\forall i\in  \mathcal{N}_h: \rho_i=\rho_{i,\min}$. As shown in Section \ref{subsec.mecSMPC}, perfect SMPC requires full utility and no additional information can be leaked except for \eqref{eq.smpcImp}, which is exactly quantified by $\rho_{i,\min}$ when $\hat{\by}=\by$.
    \item They reduce to relaxed $\epsilon$-DP when $k_i=n-1$ (i.e., $\mcalN_c= \mathcal{N}\setminus \{i\}$) and all private data are assumed to be uncorrelated. 
    That is,  if all $S_i, i\in \mathcal{N}$ are independent with each other, \eqref{eq.lbPro} becomes
    \begin{align} \label{eq.epstar}
     \rho_{i,\min}
     &=I(S_i;\{\hat{Y}_j\}_{j \in \mcalN_{c}}|\{S_j\}_{j \in \mcalN_{c}})\nonumber \\
     &+I(S_i;\{S_j\}_{j \in \mcalN_{c}}) \nonumber\\
    &=I(S_i;\{\hat{Y}_j\}_{j \in \mcalN_{c}}|\{S_j\}_{j \in \mcalN_{c}}).
\end{align}
    The above conditional mutual information is fundamentally related to DP and has been proved to be a relaxation of $\epsilon$-DP \cite{cuff2016differential}.
\end{enumerate}

\section{Example \rom{1}: Distributed average consensus}\label{sec.consensus1}
To demonstrate the benefits of the proposed metrics and the effect of the lower bound on individual privacy, we use the distributed average consensus as a canonical example. Two main reasons for choosing this problem are that it has general applicability in many signal processing tasks, such as denoising~\cite{pang2015optimal} and interpolation~\cite{narang2013signal}, and that its privacy-preserving solutions have been widely investigated in the literature \cite{li2019privacyA,gupta2017privacy,gupta2019statistical,Jane2020ICASSP,kefayati2007secure,huang2012differentially,nozari2017differentially,manitara2013privacy,mo2017privacy,he2019privacy,braca2016learning,hale2015differentially,hale2018cloud}. 

In this section, we first define the problem. After that, we introduce traditional distributed average consensus approaches and prove that they exhibit privacy leakages.  Before describing the details, we first state the following assumptions.  Let $\mathcal{N}_{i,c}=\mathcal{N}_i \cap \mathcal{N}_{c}$ and $\mathcal{N}_{i,h}=\mathcal{N}_i \cap \mathcal{N}_h$ denote the set of corrupted and honest neighbors of node $i$, respectively.
\begin{assumption}\label{asu.1}
The private data $s_i$ of each node is independent of those of the other nodes, i.e., $\forall i,j \in \mcalN, i\neq j: I(S_i;S_j)=0$. 
\end{assumption}
\begin{assumption}\label{asu.2}
The passive adversary has knowledge of the size of the network $n$, each node's neighborhood size $d_i$; and every honest node has a non-empty corrupted neighborhood, i.e., $\forall i\in \mathcal{N}_h: \mathcal{N}_{i,c} \neq \emptyset$. 
\end{assumption}

\subsection{Problem definition}
The goal of the distributed average consensus is to compute the global average of all the private data over the network, i.e.,
\begin{align}
    \by= s_{\mathrm{ave}}\bm 1,
\end{align}
where $s_{\mathrm{ave}}=n^{-1} \tsum_{i\in \mathcal{N}} s_i$. Hence, we have that $\by = n^{-1} \bm 1 \bm1^{\top}\bs$. As the nodes in the network can only communicate with the neighboring nodes, the solution is obtained iteratively.  Many approaches have been proposed to achieve this goal. Below, we introduce two types of approaches that serve as baselines for the coming section.

\subsection{Distributed linear iteration approaches}
Distributed average consensus can be obtained by applying, at every iteration $t \in \mathcal{T}$ where $\mathcal{T}=\{1,\ldots,T\}$, a linear transformation $\bW\in \mathcal{W}$ where 
\begin{align}
    \mathcal{W}=\left\{\bW \in \mathbb{R}^{n \times n} | \bW_{i j}=0 \text { if }(i,j) \notin \mathcal{E} \text { and } i \neq j\right\},
\end{align}
such that the state value $\bx$ is updated by
\[
\bx^{(t+1)}=\bW x^{(t)},
\]
and it is initialized with the private data, i.e., 
\begin{align}\label{eq.linearIni}
\bx^{(0)} =\bs.    
\end{align}

The structure of $\bW$ reflects the connectivity of the network\footnote{For simplicity, we assume that $\bW$ is constant for every iteration, which corresponds to a synchronous implementation of the algorithm. In the case of an asynchronous implementation, the transformation depends on which node will update. The results shown here are easily generalized to asynchronous systems by working with expected values.}.
In order to correctly compute the average, that is, $\bx^{(t)}\rightarrow \by = n^{-1}\bm 1 \bm 1^{\top}\bs$ as $t\rightarrow\infty$, necessary and sufficient conditions for $\bW$ are given by
\begin{itemize}
    \item[(i)] $\mathbf{1}^{\top}\bW=\mathbf{1}^{\top}$,
    \item[(ii)]  $\bW \mathbf{1}=\mathbf{1}$,
    \item[(iii)] $\alpha\left(\bW-\frac{\bm1\bm1^{\top}}{n}\right)<1$,
\end{itemize}
where $\alpha(\cdot)$ denotes the spectral radius \cite{olshevsky2009convergence}.
\vspace{.5\baselineskip}\\
\textbf{Privacy leakage:}
As the state values $x_{i}^{(t)}$ should be exchanged between nodes, based on Assumption \ref{asu.2}  we have  $X_i^{(0)} \in \mathcal{V}_i$. Thus,
\begin{align*}
   \rho_i\geq I(S_i,X_i^{(0)}) =I(S_i,S_i).
\end{align*}
We conclude that, as expected, the traditional distributed linear iteration algorithm is not privacy-preserving at all.

\subsection{Distributed optimization approaches}\label{subsec.pdmm}
Distributed average consensus can also be formed as an equivalent linear-constrained convex optimization problem given by
\begin{equation} \label{eq.setupAve}
\begin{array}{ll}{\displaystyle \min_{x_i}} & {{\displaystyle\sum_{i \in \mathcal{N}} \frac{1}{2}\|x_i - s_i\|^2_2}} \\ {\text { s.t. }} & {x_i=x_j, {\forall} (i,j)\in \mathcal{E}},\end{array}
\end{equation}
Many distributed optimizers have been proposed to solve the above problem, such as ADMM \cite{boyd2011distributed} and PDMM \cite{sherson2018derivation,zhang2018distributed}. Here, we provide an example using PDMM,  its extended augmented Lagrangian function is given by:
\begin{align}\label{eq.pdmmObj}
    \frac{1}{2}\|\bx - \bs\|^2_2 + (\bP\blambda^{(t)})^T\!\!\bC \bx + \frac{c}{2}\|\bC \bx+ \bPC \bx^{(t)}\|_2^2,
\end{align}
and the updating equations are 
\begin{align}
\bx^{(t+1)} &= \left(\bm I + c\bC^{\top}\bC\right)^{-1}\left(\bs - c\bC^{\top}\bP\bC\bx^{(t)} - \bC^{\top}\bP\blambda^{(t)}\right),\label{eq.xUp} \\
\blambda^{(t+1)} &= \bP\blambda^{(t)} + c(\bC \bx^{(t+1)} + \bPC \bx^{(t)}), \label{eq.lamdaUp}
\end{align}
where $c>0$ is a constant for controlling the convergence rate. $\blambda \in \mathbb{R}^{2 m}$ denotes the introduced dual variable and matrix $\bC \in \mathbb{R}^{2 m \times n}$ is related to the graph incidence matrix $\bB$. Let the subscript $i|j$ be a directed identifier that denotes the directed edge from node $i$ to $j$ and $i,j$ be an undirected identifier. In PDMM, each edge  $e_l=(i,j)\in\mathcal{E}$ corresponds two dual variables: $\lambda_{l}=\lambda_{i|j}$, $\lambda_{l+m}=\lambda_{j|i}$ and two rows in matrix $\bC$: $\bC_{li}=\bB_{i|j}=1$, $\bC_{(l+m)j}=\bB_{j|i}=-1$ if and only if $i<j$.  Of note, $\bP \in \mathbb{R}^{2 m \times 2 m}$ denotes a symmetric permutation matrix which flips the upper $m$ with the lower $m$ rows of the matrix it applies. Thus, $\forall (i,j)\in \mathcal{E}: \lambda_{j|i} = \left(\bP\blambda\right)_{i|j}$ and $\bC + \bPC = [\bB^{\top} \, \bB^{\top}]^{\top}$.

The local updating functions for each node become
\begin{align}
&x_i^{(t+1)} = \frac{s_{i} + \sum_{j\in \mathcal{N}_{i}} \left(cx_j^{(t)} - \bB_{i|j}\lambda_{j|i}^{(t)} \right)}{1+cd_i}, \label{eq.xiup} \\
 &\lambda_{i|j}^{(t+1)} = \lambda_{j|i}^{(t)} + c\big(\bB_{i|j}x_{i}^{(t+1)}+\bB_{j|i}x_{j}^{(t)} \big). \label{eq.lamdaiup}
\end{align}
$\bx^{(t)}$ has been proven to converge geometrically (linearly on a logarithmic scale) to optimum $\bx^* =s_{\mathrm{ave}}\bm 1$, given arbitrary initialization of both $\bx$ and  $\blambda$. \\
\textbf{Privacy leakage:} 
By inspecting \eqref{eq.xiup} we can see that the privacy leakage about $s_i$ depends not only on $\bx$ but also on $\blambda$. It is thus important also to analyze the convergence behavior of the dual variable $\blambda$.  We first consider two successive $\blambda$-update in \eqref{eq.lamdaUp} given by
\begin{align}\label{eq.pdmmLamUp}
\blambda^{(t+2)} = \blambda^{(t)} + c(\bC\bx^{(t+2)} + 2\bPC\bx^{(t+1)} + \bC\bx^{(t)}),
\end{align}
as $\bP^2 = \bm I$. 
Let $\bar{H}= \operatorname{span}(\bC)+\operatorname{span}(\bPC)$ and $\bar{H}^{\perp} = \operatorname{null}(\bC^{\top}) \cap \operatorname{null}((\bPC)^{\top})$. We can see that every two $\blambda$-updates affect only $\Pi_{\bar{H}} \blambda \in \bar{H}$ where $\Pi_{\bar{H}}$ denotes the orthogonal projection onto $\bar{H}$. It is proven that if $\blambda^{(0)}\in \bar{H}$, the dual variable will be ensured to converge to an optimum $\blambda^{*}$ \cite{sherson2018derivation} given by
\begin{equation}\label{eq.lamdaOptimum}
\blambda^* = -\left( \!\! \begin{array}{c} \bC^{\top} \\ (\bPC)^{\top}\end{array} \!\! \right)^{\!\!\dagger} \left( \!\! \begin{array}{c} \bx^*-\bs  + c\bC^{\top}\!\bC\bx^* \\ \bx^*-\bs + c \bC^{\top}\!\bPC\bx^* \end{array} \!\! \right) + c\bC\bx^*.
\end{equation}
Note that the traditional distributed optimizer generally initializes $\blambda^{(0)}\in \bar{H}$ to ensure that $\blambda\rightarrow \blambda^{*}$. To do so, zero initialization is the simplest way as it does not require any coordination between nodes. In addition, 
zero initialization of both $\bx$ and  $\blambda$ 
give the smallest initial error resulting in the smallest number
of iterations to converge. As a consequence, by inspecting \eqref{eq.xiup} we have 
\begin{align}
&x_i^{(1)} =  \frac{s_{i}}{1+cd_i}.
\end{align}
As the constant $c$ is globally known to all nodes and the neighborhood size $d_i$ is known to the adversary based on Assumption \ref{asu.2}, the private data $s_{i}$ can be reconstructed by the adversary. We thus have
\begin{align*}
    \rho_i\geq I(S_i,X_i^{(1)})=I(S_i,S_i),
\end{align*}
as $X_i^{(1)} \in \mathcal{V}_i$.
Hence, we conclude that traditional distributed optimization algorithms leak private information.

\section{Example \rom{2}: Privacy-preserving distributed average consensus}\label{sec.consensus2}
In the previous section, we showed that information about the private data is revealed during the data exchange step. As a consequence, one way to protect privacy is to not exchange private data directly, but to insert noise to obtain an obfuscated version and then exchange the obfuscated data with other nodes. In what follows, we will first present an information-theoretic result regarding using noise insertion to achieve privacy-preservation. After that, we will introduce existing privacy-preserving distributed average consensus approaches and quantify their performances using the proposed metrics.  
\subsection{Noise insertion for privacy preservation}\label{subsec.noiseIns}
\begin{proposition}\label{prop.2} (Arbitrary small information loss can be achieved through noise insertion.) Let private data $s$ and inserted noise $r$ denote a realization of independent random variable $S$ and $R$ with variance $\sigma^{2}_{S},\sigma^{2}_{R}<\infty$, respectively. Let $Z=S+R$.  Given arbitrary small $\epsilon\in \mathbb{R}_{>0}$, there exists $\sigma^{2}_{R}$ that satisfies 
\begin{align}
    I(S;Z)\leq \epsilon.
\end{align}
In addition,  $\sigma^{2}_{R}$ is bounded by 
\begin{align}\label{eq.sigmaR}
    \sigma^{2}_{R}\geq \frac{\sigma^{2}_{S}}{2^{2\epsilon}-1},
\end{align}
if we choose to insert Gaussian noise.
\begin{proof}
See Appendix~\ref{pf.prop2}.
\end{proof}
\end{proposition}
Based on the design of the noise insertion process,  we broadly classified existing approaches into two classes: zero-sum noise insertion and subspace noise insertion. We first introduce the former case,
the main idea of zero-sum noise insertion comes from the nature of the distributed average consensus. Let $r_i$ denote the noise added by node $i$; the estimated output is thus given by
\begin{align}\label{eq.hatyi}
    \hat{y}_i&=\frac{1}{n}\sum_{i\in \mcalN} (s_i+ r_i) =y_i+\frac{1}{n}\sum_{i\in \mcalN}r_i.
\end{align}
Clearly, if the sum of all inserted noise is zero, full output utility will be achieved as $\forall i\in \mcalN,~ \hat{y}_i=y_i$. Now we will proceed to introduce two different approaches, including DP and SMPC,  which aim to insert zero-sum noise in a distributed manner. 
\subsection{Statistical zero-sum noise insertion using DP}
DP-based approaches~\cite{kefayati2007secure,huang2012differentially,nozari2017differentially} mostly apply zero-mean noise insertion to achieve zero-sum in a statistical sense. Variants exist in designing the noise insertion process, here we give one simple example to illustrate the main idea: each node $i$ initialize its state value by adding zero-mean noise $r_i$ to its private data. That is, the state value initialization \eqref{eq.linearIni} is replaced with
\begin{align} \label{eq.x0SS}
\forall i\in \mathcal{N}: x_i^{(0)}&=s_i+r_i,
\end{align}
and then arbitrary distributed averaging algorithms (e.g., linear iterations or distributed optimization) can be adopted to compute the average.   
\\
\subsubsection{Output utility analysis}
Assume that all inserted noise are independent and identically distributed random variables with zero-mean and variance $\sigma^2$. 
Denote $r=\sum_{i\in \mcalN}r_i$ and $\bar{r}=\frac{r}{n}$ as the sum of all inserted noise and its average, respectively; thus, $R$ and $\bar{R}$ are also zero-mean, and their variances are $n\sigma^2$ and $\frac{\sigma^2}{n}$, respectively. 
Based on \eqref{eq.hatyi} the output utility of node $i$ is 
\begin{align}\label{eq.uiDP}
    \forall i\in \mcalN:~ u_i=I(Y_i;Y_i+\bar{R}).
\end{align}
 \subsubsection{Lower bound analysis}
As mentioned in Section \ref{subsec.dp}, DP assumes $n-1$ corrupted nodes implying $\mathcal{N}_{c}=\{j\}_{j\in \mcalN,j\neq i}$.
With Assumption \ref{asu.1} the lower bound on individual privacy reduces to \eqref{eq.epstar}. Therefore,
\begin{align}
   \rho_{i,\min}&=I(S_i;Y_i+\bar{R}|\{S_j\}_{j\in \mathcal{N}_{c}})\nonumber\\
    &\stackrel{{\text{(a)}}}{=} I(S_i;\tsum_{j\in \mathcal{N}}S_j+R|\{S_j\}_{j\in \mathcal{N}_{c}})\nonumber\\
    &\stackrel{{\text{(b)}}}{=} I(S_i;\tsum_{j\in \mathcal{N}}S_j+R,\{S_j\}_{j\in \mathcal{N}_{c}})-I(S_i;\{S_j\}_{j\in \mathcal{N}_{c}})\nonumber\\
     &\stackrel{{\text{(c)}}}{=} I(S_i;S_i+R,\{S_j\}_{j\in \mathcal{N}_{c}})\nonumber\\
    &\stackrel{{\text{(d)}}}{=}I(S_i;S_i+R) \label{eq.lowerDP},
\end{align}
where (a) comes from Assumption \ref{asu.2} that $n$ is known to the adversary; (b) comes from the definition of conditional mutual information; (c) holds as $I(S_i;\{S_j\}_{j\in \mathcal{N}_{c}})=0$ from Assumption \ref{asu.1} and the fact  $S_i+R,\{S_j\}_{j\in \mathcal{N}_{c}}$ is a sufficient statistic of $\tsum_{j\in \mathcal{N}}S_j+R,\{S_j\}_{j\in \mathcal{N}_{c}}$; (d) holds from the fact that $\{S_j\}_{j\in \mathcal{N}_{c}}$ is independent of $S_i+R$.
\subsubsection{Individual privacy analysis}
Denote vector $X^{(t)}=[X_1^{(t)},\ldots,X_n^{(t)}]^{\top}$, 
with $n-1$ corrupted nodes  all the information seen by the adversary over the algorithm is 
\begin{align}
    \mathcal{V}_i=\{\hat{Y}_j,S_j,R_j,X^{(t)}\}_{j\in \mathcal{N}_{c},t\in \mathcal{T}},
\end{align}
where $\hat{Y}_i=X_i^{(T)}$. 
We can see that computing $I(S_i;\mathcal{V}_i)$ requires to analyze the information flow over the whole iterative process, this imposes challenges as keeping track of information loss throughout all iterations is difficult. We, therefore, simplify the privacy analysis through the following result.
\begin{remark}\label{rm.1} (Information release after the initialization step does not leak additional information.) For all iterations $t\geq 1$, the sequence 
$S_i\rightarrow X^{(0)}\rightarrow  X^{(t)}$ forms a Markov chain; on the basis of the data processing inequality \cite{cover2012elements}, we have
\begin{align}\label{eq.dpIneq}
    \forall t\geq 1: I(S_i;X^{(0)})\geq I(S_i;X^{(t)}).
\end{align} 
We conclude that analyzing the individual privacy by using the information flow in the initialization is sufficient, i.e., 
\begin{align}
    I(S_i;X^{(0)})=I(S_i;X^{(0)},X^{(1)},\ldots, X^{(T)}).
\end{align} 
\end{remark}
Given the above Remark, we have 
\begin{align}
    I(S_i;\mathcal{V}_i)=&I(S_i;\{S_j,R_j,X^{(0)}\}_{j\in \mathcal{N}_{c}})\nonumber\\
    =&I(S_i;X_i^{(0)}|\{S_j,R_j,X_j^{(0)}\}_{j\in \mathcal{N}_{c}})\nonumber\\
    &+I(S_i;\{S_j,R_j,X_j^{(0)}\}_{j\in \mathcal{N}_{c}})\nonumber\\
    =&I(S_i;X_i^{(0)}),
\end{align}
where the last equality holds, as $\{S_j,R_j,X_j^{(0)}\}_{j\in \mathcal{N}_{c}}$ is independent of both $S_i$ and $X_i^{(0)}$.
The individual privacy thus becomes
\begin{align}\label{eq.priDP}
    \rho_i=I(S_i;S_i+R_i).
\end{align}
In conclusion, with the proposed metrics DP based approaches achieve
\begin{align*}
    \left(I(Y_i;Y_i+\bar{R}),I(S_i;S_i+R_i)\geq I(S_i;S_i+R),n-1\right).
\end{align*}
By inspecting the above result, we have the following remark.
\begin{remark}\label{rm.2} (In the distributed average consensus, DP always has a trade-off between the output utility and individual privacy.)
As both output utility \eqref{eq.uiDP} and individual privacy \eqref{eq.priDP} are dependent on the inserted noise, with Proposition \ref{prop.2} we describe two extreme cases based on the variance of inserted noise:
\begin{align}
&\sigma^2\rightarrow \infty \Rightarrow   u_i =0, \rho_i=0,\\
&\sigma^2=0\Rightarrow u_i =I(Y_i;Y_i), \rho_i=I(S_i;S_i).
\end{align}
\end{remark}
Hence DP has a trade-off between privacy and utility. 
Of note, the conclusion that DP based approaches can not achieve full utility has been shown in \cite{nozari2017differentially}; here, we provide a simpler proof in terms of mutual information.

\subsection{Exact zero-sum noise insertion using SMPC}
Unlike the DP based approaches, which have a privacy-utility trade-off, the SMPC based approaches have a feature of ensuring full utility without compromising privacy.  However, there is no “free lunch”; the price is that the robustness over $n-1$ corrupted nodes is no longer achievable. Existing SMPC based approaches \cite{gupta2017privacy,gupta2019statistical,li2019privacyA} have applied additive secret sharing \cite{Cramer2015} to construct exact zero-sum noise through coordinated noise insertion. To do so, each node $i$ first sends each neighbor $j\in \mcalN_i$ a random number $r_i^{j}$ and receives a random number $r_j^{i}$ from each of its neighbors. After that node $i$ constructs its noise by 
\begin{align} 
 r_{i}=\sum_{j\in \mcalN_i} r_{i|j}, \label{eq.riSS}
\end{align}
where 
\begin{align}
    r_{i|j}=r_j^{i}-r_i^{j}\label{eq.rijSS}.
\end{align}
Of note, all the random numbers $\{r_i^{j}\}_{(i,j)\in \mathcal{E}}$ are independent of each other. 
After constructing noise, similar as DP based approaches, each node initializes its state value using  \eqref{eq.x0SS}
and then arbitrary distributed averaging algorithm can be adopted. 
\subsubsection{Output utility analysis}
In SMPC the noise is constructed such that they all sum to zero: 
\begin{align}\label{eq.zerosSMPC}
    \sum_{i\in \mcalN}r_i =\sum_{(i,j)\in \mathcal{E}} r_{i|j}=0,
\end{align}
as $r_{i|j}=-r_{j|i}$. Full utility is thus obtained as $\hat{y}_i=y_i$:
\begin{align}
    \forall i\in \mcalN:~ u_i=I(Y_i;Y_i).
\end{align}
\subsubsection{Lower bound analysis}
With full utility, the lower bound \eqref{eq.epstar} becomes
\begin{align}\label{eq.epsstarSMPC}
  \rho_{i,\min} &= I(S_i;\{Y_j\}_{j\in \mathcal{N}_{c}}|\{S_j\}_{j\in \mathcal{N}_{c}})\nonumber\\
    &\stackrel{{\text{(a)}}}{=}I(S_i;\sum_{j\in \mathcal{N}_{h}} S_j),
\end{align}
where (a) holds on the basis of Assumption \ref{asu.1} and \ref{asu.2}.
\subsubsection{Individual privacy analysis}
Let $\mathcal{E}_c=\{(i,j) \in \mathcal{E}, (i,j)\notin \mathcal{N}_h\times \mathcal{N}_h\}$ denote the set of corrupted edges. 
For arbitrary honest node $i\in \mathcal{N}_h$, all information that the adversary sees through the algorithm is given by
\begin{align*}
    \mathcal{V}_i=\{\{S_j\}_{j\in \mathcal{N}_c}, \tsum_{j\in \mathcal{N}}S_j,\{R_i^j\}_{(i,j) \in\mathcal{E}_c}, \{X^{(t)}\}_{t\in \mathcal{T}}\},
\end{align*}
where $\tsum_{j\in \mathcal{N}}s_j=ns_{\mathrm{ave}}$ is known as the adversary knows both $n$ and correct average $s_{\mathrm{ave}}$; full vector $X^{(t)}$ is known because of Assumption \ref{asu.2}.
Let $\mathcal{G}_h'$ denote the component (i.e., connected subgraph) consisting of node $i$ after removal of all corrupted nodes; its node set is denoted by $\mathcal{N}_h'\subseteq \mathcal{N}_h$. We have the following result which simplifies the individual privacy analysis.
\begin{proposition}\label{prop.3} Given $\mathcal{V}_i$, information flow within the subgraph  $\mathcal{G}_h'$ provides a sufficient statistic for inferring $S_i$. More specifically, we have
\begin{align}
  \forall i\in \mathcal{N}_h': I(S_i;\mathcal{V}_i)=I(S_i;\{S_j+\tsum_{k\in \mathcal{N}_{j, h}}R_{j|k}\}_{j\in \mathcal{N}_h'}).   
\end{align}
\begin{proof}
See Appendix~\ref{pf.prop3}. 
\end{proof}
\end{proposition}
With the knowledge of $\{S_j+\tsum_{k\in \mathcal{N}_{j, h}}R_{j|k}\}_{j\in \mathcal{N}_h'}$, the adversary has different ways to infer information about $S_i$ for example by looking at (1) the term $I(S_i;S_i+\tsum_{j\in \mathcal{N}_{i, h}}R_{i|j})$ in which we can see that node $i$ should have at least one honest neighbor, i.e., $\mathcal{N}_{i, h}\neq\emptyset$, otherwise $S_i$ will be revealed; therefore, among the neighboring nodes the maximum number of corrupted nodes can be tolerated is $k_i= d_i-1$; (2) the partial sum of subgraph $\mathcal{G}_h'$: $I(S_i;\tsum_{j\in \mathcal{N}_h'}S_j)$ since 
\begin{align}
    \sum_{j\in \mathcal{N}_h'}S_j=\sum_{j\in \mathcal{N}_h'}(S_j+\tsum_{k\in \mathcal{N}_{j, h}}R_{j|k}),
\end{align} 
as $R_{j|k}=-R_{k|j}$ from \eqref{eq.rijSS}.  Since this partial sum can always be determined regardless of the amount of noise insertion, we then have  
\begin{align}
    I(S_i;\mathcal{V}_i)\geq I(S_i;\tsum_{j\in \mathcal{N}_h'}S_j).
\end{align}
Due to Proposition \ref{prop.2} we conclude that the minimum information loss $I(S_i;\tsum_{j\in \mathcal{N}_h'}S_j)$ can be achieved through noise insertion. The individual privacy is thus given by 
\begin{align}\label{eq.smpcind}
\rho_i=I(S_i;\tsum_{j\in \mathcal{N}_h'}S_j).
\end{align}
In conclusion, with the proposed metrics SMPC based approaches achieve 
\begin{align}\label{eq.smpcAll}
    \left(I(Y_i;Y_i),I(S_i;\tsum_{j\in \mathcal{N}_h'}S_j)\geq I(S_i;\tsum_{j\in \mathcal{N}_h}S_j),d_i-1\right).
\end{align}
\begin{remark} \label{rm.fullPri} (Conditions for achieving perfect individual privacy and full output utility using the SMPC based approaches in the distributed average consensus.) Given Definition \ref{def.1}, by inspecting \eqref{eq.smpcAll} we conclude that the SMPC based approaches is able to achieve both full output utility and perfect individual privacy if $\forall i\in \mcalN_h: \mathcal{N}_h'=\mcalN_h$ and $|\mcalN_h|\geq 2$, i.e., the graph is still connected after removal of all corrupted nodes.
\end{remark}

The main limitation of the above zero-sum noise insertion approaches is that it is hard to be generalized to problems other than distributed average consensus. To mitigate this problem, subspace noise insertion approach proposed to exploit the graph structure of distributed signal processing. Below, we introduce a recently proposed  approach called distributed optimization based subspace perturbation (DOSP).
\subsection{Subspace noise insertion using DOSP}
The DOSP approach \cite{Jane2020ICASSP,Jane2020Arxiv} is distinct from both the DP and SMPC based approaches, because it can ensure full utility without compromising privacy and does not require coordinated noise insertion.  In particular, DOSP does not introduce additional variables for noise insertion but exploit the dual variable to construct the noise. By inspecting \eqref{eq.xiup}, the noise for each node $i$ is constructed as
\begin{align}
   \forall t\in \mathcal{T}: r_i^{(t)} =\sum_{j\in \mathcal{N}_{i}} (\bB_{i|j}\lambda_{j|i}^{(t)}),
\end{align}
in which the dual variables of the corrupted neighbors, i.e., $\{\lambda_{j|i}^{(t)}\}_{j\in \mathcal{N}_{i, c}}$ are known to the adversary. 
As shown in \cite{sherson2018derivation}, the dual variable $\blambda^{(t)}$ composites of two parts: the convergent component $\Pi_{\bar{H}}\blambda^{(t)}\rightarrow \blambda^*$ and the non-convergent component $(\bI-\Pi_{\bar{H}})\blambda^{(t)}=\bP^{t}\left(\bI-\Pi_{\bar{H}}\right) \blambda^{(0)}$ where $\bP^{2}=\bm I$. Therefore, we have 
\begin{align}
\sum_{j\in \mathcal{N}_{i,h}} (\bB_{i|j}\lambda_{j|i}^{(t)})=&\sum_{j\in \mathcal{N}_{i,h}} (\bB_{i|j}(\Pi_{\bar{H}}\blambda^{(t)})_{j|i}) \nonumber \\
+&\sum_{j\in \mathcal{N}_{i,h}} \left(\bB_{i|j}(\bP^{t}(\bI-\Pi_{\bar{H}}) \blambda^{(0)})_{j|i}\right)    
\end{align}
The main idea of subspace noise insertion is to exploit the non-convergent component of the dual variables as subspace noise for guaranteeing the privacy. That is,  $\sum_{j\in \mathcal{N}_{i,h}} \left(\bB_{i|j}(\bP^{t}(\bI-\Pi_{\bar{H}}) \blambda^{(0)})_{j|i}\right)$ is to protect the  private data $s_i$ of honest node $i$ from being revealed to others. Because it only depends on the initialization and thus its variance can be made arbitrarily large to fit with different privacy levels on the basis of Proposition \ref{prop.2}.


Before discussing how to implement the subspace noise, we first state the following remark.
\begin{remark}(There is always a non-empty subspace for noise insertion as long as $m\geq n$.) Since $[\bC~ \bPC]\in \mathbb{R}^{2m\times2n}$ can be viewed as a new graph incidence matrix with $2n$ nodes and $2m$ edges \cite{Jane2020Arxiv}, we thus have $\mathrm{dim}(\bar{H}) \leq 2n-1$, and $\bar{H}^\perp$ is non-empty if $m\geq n$.
\end{remark}
In DOSP,  each node only needs to randomly initialize its own dual variables $\{\lambda_{i|j}^{(0)}\}_{j\in \mathcal{N}_i}$; we thus have non-zero subspace noise  $(\bI-\Pi_{\bar{H}}) \blambda^{(0)}\neq \bm 0$ with probability 1 as long as $m\geq n$. Hence, DOSP does not require any coordination between nodes for noise construction. 
\subsubsection{Output utility analysis}
Beyond not requiring coordination between nodes, DOSP also ensures full output utility regardless of the amount of inserted noise \cite{Jane2020Arxiv}, because the updating of the optimization variable $\bx$ is perpendicular to subspace $\bar{H}^\perp$ by inspecting \eqref{eq.pdmmObj}, i.e.,
\begin{align}
   \big(\left(\bI-\Pi_{\bar{H}}\right)\blambda^{(t)}\big)^{\top}\bC \bx= 0.
\end{align}
Full output utility is thus achieved.
\subsubsection{Lower bound analysis}
As full output utility is achieved, the lower bound on DOSP is the same as \eqref{eq.lowerDP} in the SMPC based approach.
\subsubsection{Individual privacy analysis}
The information collected by the adversary throughout the whole algorithm is given by
\begin{align}
    \mathcal{V}_i=\{\{S_j\}_{j\in \mathcal{N}_c}, \tsum_{j\in \mathcal{N}}S_j,\{\Lambda_{i|j}^{(t)},X^{(t)}\}_{(i,j) \in\mathcal{E}_c,t\in \mathcal{T}}\}.
\end{align}
We have the following result which simplifies the privacy analysis:
\begin{align}\label{eq.dospind}
   I(S_i;\mathcal{V}_i)= &I(S_i;\{S_j-{\tsum_{k\in \mathcal{N}_{j, h}}} \bB_{j|k}\Lambda^{(t)}_{k|j}\}_{j\in \mathcal{N}_h,t=0,1} \nonumber\\
   &|\{S_j\}_{j\in \mathcal{N}_c},\{\Lambda_{i|j}^{(0)}\}_{(i,j), \in\mathcal{E}_c}),
\end{align}
where the proof is shown in Appendix \ref{pf.dospind}.

We note that, similarly to the above SMPC based approach,  the partial sum in subgraph $\mathcal{G}_h'$ can also be computed by the adversary. In fact, the partial sum can be divided into two parts: 
\begin{align}
  \tsum_{j\in \mathcal{N}_h'}S_j &= \tsum_{j\in \mathcal{N}_h',t=0,1}\big(S_j-{\tsum_{k\in \mathcal{N}_{j, h}}} \bB_{j|k}\Lambda^{(t)}_{k|j}\big) \nonumber\\
    &+\tsum_{(j,k)\in \mathcal{E} \cap \mathcal{N}_h' \times \mathcal{N}_h' ,t=0,1} \bB_{j|k}\Lambda^{(t)}_{k|j}.
\end{align}
The first term is given by \eqref{eq.dospind}.  The second term can also be determined by using \eqref{eq.lamdaiup} and the fact that $
\bB_{i|j} \lambda_{j|i}^{(0)}+\bB_{j|i} \lambda_{i|j}^{(1)}=\bB_{i|j} (\lambda_{j|i}^{(0)}- \lambda_{i|j}^{(1)})$. Therefore, the partial sum $\tsum_{j\in \mathcal{N}_h'}S_j$ can be computed by the adversary. 

As the partial sum can be computed, the rest analysis follows a similar line as the above SMPC based approaches and we conclude that with the proposed metrics, DOSP also achieves \eqref{eq.smpcAll}. In addition, Remark \ref{rm.fullPri} also holds for DOSP.

\section{Comparisons, numerical results, and discussion}\label{sec.numRes}
In this section, we first compare all the above discussed approaches and then demonstrate their numerical results. Finally, we will discuss on principles for designing algorithms.
\begin{table*}[ht!]
\begin{center}
\centering \caption{Comparisons of existing information-theoretic solutions for the distributed average consensus} \label{table.1}
{
\begin{tabular}{|c|c|c|c|}
\hline  & DP 
& SMPC 
& DOSP  \\
\hline
Adversary models&\multicolumn{3}{c|}{Passive, Eavesdropping}   \\
\hline
Coordinated noise insertion &No&Yes&No\\
\hline
Output utility&$u_i={I(Y_i;Y_i+\bar{R})}$&\multicolumn{2}{c|}{$u_i=I(Y_i;Y_i)$}\\
 \hline
Individual privacy&$\rho_i={{I(S_i;S_i+R_i^{(0)})}}$
&\multicolumn{2}{c|}{$\rho_i={I(S_i;\sum_{j\in \mathcal{N}_h'} S_j)}$}\\
 \hline
Lower bound on individual privacy&$\rho_{i,\min}={I(S_i;S_i+R)}$&\multicolumn{2}{c|}{$\rho_{i,\min}={I(S_i;\sum_{j\in \mathcal{N}_h} S_j)}$}\\
 \hline
Maximum number of corrupted nodes &$k_i=n-1$ out of $n$&\multicolumn{2}{c|}{$k_i=d_i-1$ out of $d_i$} \\
 \hline
Channel encryption cost& 0 &$1$ &$1$\\
 \hline
\end{tabular}}
\end{center}
\vskip -16pt
\end{table*}

\subsection{Comparisons of existing approaches} \label{subsec.dpSmpc}
In Table \ref{table.1}, we compare the discussed approaches in terms of several important parameters.  Firstly, we can see that in the context of distributed average consensus, SMPC and DOSP achieve exactly the same performance, except the fact that SMPC requires coordination between nodes to construct the sum of noise to zero (i.e., the steps required in \eqref{eq.riSS}). Secondly, compared to SMPC and DOSP, DP is robust against $n-1$ corrupted nodes, but it suffers from a privacy-utility trade-off. Thirdly, for SMPC and DOSP,  $k_i=d_i-1$ is only dependent on neighborhood size $d_i$ but not on the whole network size. If the graph is fully connected, it reduces to $k_i=n-2$.    Finally, when dealing with an eavesdropping adversary, DP is the most lightweight as it protects privacy even though all transmitted messages are eavesdropped; while, both SMPC and DOSP require securely encrypted channels at the initialization step only to guarantee the noise $\{r_i\}_{i\in \mathcal{N}_h}$ is not revealed to the adversary. 

\subsection{Numerical results}
\begin{figure}
    \centering
    \includegraphics[width=.42\textwidth]{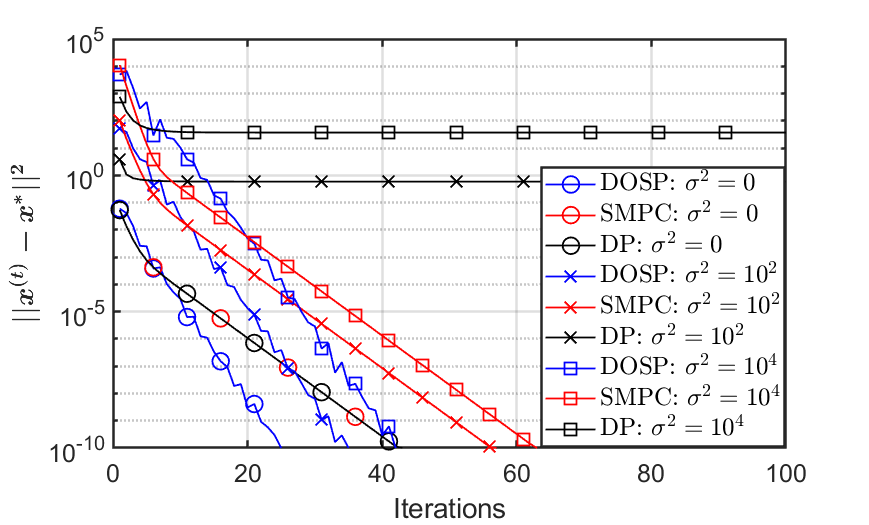}
    \caption{Convergence behaviors of DOSP, SMPC and DP based approaches under three different amounts of noise insertion.}
    \label{fig.conAve}
\end{figure}
\begin{figure}
    \centering
    \includegraphics[width=.42\textwidth]{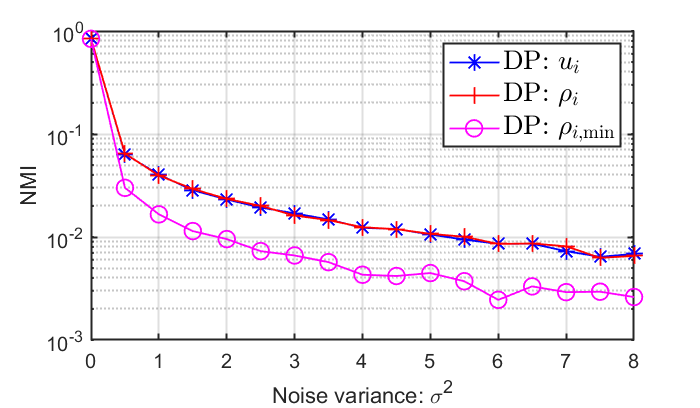}
    \caption{Normalized mutual information of output utility, individual privacy, and its lower bound for arbitrary honest node $i$ in terms of the amount of noise insertion by using the DP based approach.}
    \label{fig.trdoff}
\end{figure}
\begin{figure*}[ht]
\begin{subfigure}{0.24\textwidth}
\includegraphics[width=.9\textwidth]{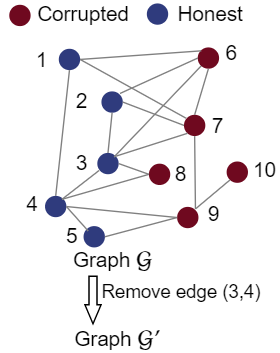} 
\caption{}
\end{subfigure}
\begin{subfigure}{0.37\textwidth}
\includegraphics[width=.99\textwidth]{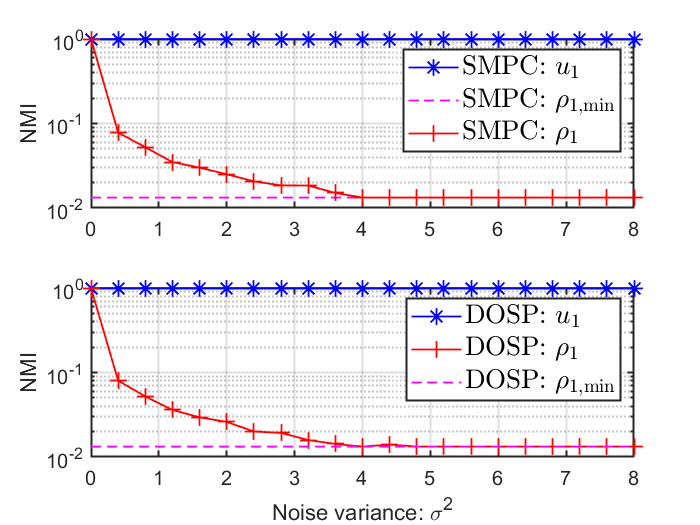}
\caption{}
\end{subfigure}
\begin{subfigure}{0.37\textwidth}
\includegraphics[width=.99\textwidth]{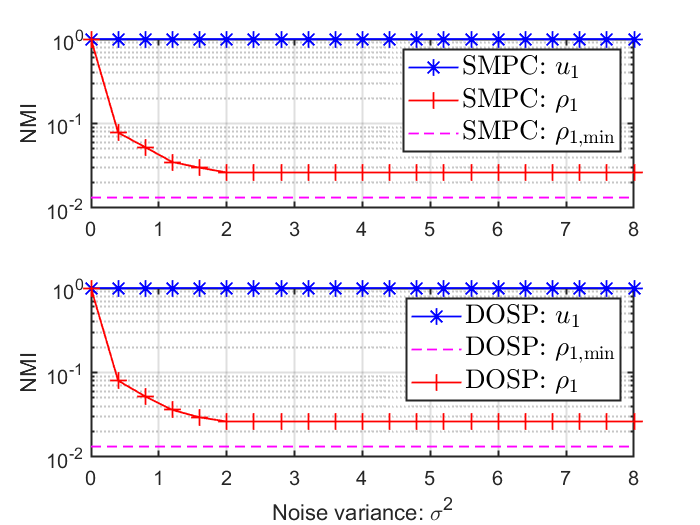}
\caption{}
\end{subfigure}
\caption{(a) Two sample graphs in which $\mathcal{G}'$ and $\mathcal{G}$ differ in only one edge. Normalized mutual information of output utility,  individual privacy, and its lower bound for honest node $1$ in terms of the amount of noise insertion by using SMPC and DOSP approaches under (b) graph $ \mathcal{G}$ and (c) graph $\mathcal{G}'$.}
\label{fig.perf}
\end{figure*}

The numerical results include two parts: the convergence behavior analysis, and visualization of both utility and privacy in terms of mutual information. To do so, we simulated a geometrical graph with $n=10$ nodes, and set the radius as $r^2 = 2\frac{\log n}{n}$ to ensure a connected graph with high probability \cite{dall2002random}. For simplicity, all private data have a Gaussian distribution with unit variance, and all the noise used in the DP, SMPC and DOSP approaches also follows a Gaussian distribution with variance $\sigma^2$.  
\subsubsection{Convergence behavior}
In Fig. \ref{fig.conAve} we present the convergence behaviors of existing algorithms under different amounts of noise insertion, i.e., different noise variances. There, we can see that all algorithms can achieve the correct average if there is no noise, i.e., $\sigma^2=0$; the DOSP and SMPC based approaches ensure the correct average regardless of the amount of inserted noise, whereas the accuracy of the DP based approach is compromised by increasing the amount of noise insertion.  
\subsubsection{Visualization of both utility and privacy}
To validate the information-theoretic results, i.e., output utility, individual privacy, and its lower bound, presented in Table \ref{table.1}, we ran $10^{4}$ Monte Carlo simulations and used the non-parametric entropy estimation toolbox (npeet) \cite{ver2000non} to estimate the normalized mutual information (NMI). 
\\ \textbf{Privacy and utility results of the DP based approach.}
As shown in Fig. \ref{fig.trdoff}, we can see that lower individual privacy can be achieved by increasing the noise variance; however, the output utility will be deteriorated. Hence, Remark \ref{rm.2} is validated that there is a trade-off between privacy and utility using the DP based approaches. Additionally, it also shows that the DP based approaches can not reach perfect individual privacy as long as there is noise insertion, i.e., $\sigma^2>0$. 
\\ \textbf{Privacy and utility results of the DOSP and SMPC based approaches.}
Unlike the DP based approaches, which consider only the case of $n-1$ corrupted nodes, the performances of SMPC and DOSP are dependent on the number of corrupted nodes in the neighborhood and the graph topology. To demonstrate this effect, in Fig. \ref{fig.perf} (a), we simulated two graphs satisfying Assumption \ref{asu.2}; i.e., every honest node is connected to at least one corrupted node. The privacy-utility results of the DOSP and SMPC based approaches under these two graphs are shown in Fig. \ref{fig.perf} (b) and (c) respectively. We validate the following theoretical results regarding utility and privacy: 
\begin{itemize}
    \item SMPC and DOSP both ensure full utility regardless of the amount of  noise, thus of the privacy level;
    \item their optimum individual privacy is only related to the partial sum in subgraph $\mathcal{G}_h'$, the connected component consists of node $1$ after removal of all corrupted nodes,  after ensuring the variance of inserted noise is sufficiently large;
    \item as expected, under graph $\mathcal{G}$ they are able to obtain perfect individual privacy, i.e., Remark \ref{rm.fullPri} is validated. 
\end{itemize}

\subsection{Discussion of principles of algorithm designs}\label{subsec.dis}
We now provide some implications on how to design appropriate algorithms for different applications. A typical way to design privacy-preserving solutions is to use the off-the-shelf tools such as DP, SMPC and DOSP. It is thus important to know their relations before designing solutions. We have the following result. 
\begin{remark}\label{rm.muExc}
(DP and SMPC/DOSP are mutually exclusive for applications satisfying $I(S_i;\{S_j,Y_j\}_{ j\in \mcalN,j\neq i })=I(S_i;S_i)$.) The reason is that if DP achieves full utility $\hat{\by}=\by$ like SMPC or DOSP, it is not privacy-preserving at all since  $\rho_{i, \min}=I(S_i;S_i)$.
\end{remark}
As a consequence, for applications like distributed average consensus satisfying $I(S_i;\{S_j,Y_j\}_{ j\in \mcalN,j\neq i })=I(S_i;S_i)$, we have the following suggestions for algorithm designs: 
\begin{enumerate}
   \item If the application requires the algorithm robustness that each node does not trust any other node in the network, i.e., $n-1$ corrupted nodes, then adopt DP based approaches; be aware that there is a trade-off between privacy and utility.
    \item If the application has very high requirements for the accuracy of function output, e.g., full utility, then both SMPC and DOSP are options, but it can not be robust to $n-1$ corrupted nodes.
\end{enumerate}
For the rest of applications, first compute the lower bound $\rho_{i, \min}=I(S_i;\{S_j,Y_j\}_{ j\in \mcalN_c })$ under the condition of obtaining full utility. After that, 
\begin{enumerate}
    \item if $\rho_{i, \min}$ is tolerable, then use either SMPC or DOSP to realize both perfect individual privacy and full output utility (might be dependent on the graph topology);
    \item if $\rho_{i, \min}$ is not tolerable, one option is to combine SMPC or DOSP with DP to decrease the lower bound by compromising the output utility. 
\end{enumerate}

\section{Conclusions}\label{sec.conclu}
In this paper, we first proposed information-theoretic metrics for quantifying the algorithm performance in terms of output utility, individual privacy and algorithm robustness. The proposed metrics are general and can reduce to well-known frameworks including SMPC and DP under certain conditions.  Then we derived several theoretical results in terms of mutual information.  In particular, the lower bound on individual privacy indicates the best privacy level can possibly be achieved before designing algorithms.
Moreover,  with a concrete example we explicitly analyzed, compared and related the state-of-the-art algorithms including DP, SMPC and DOSP. Furthermore, we also visualized all the theoretical results with numerical validations. 

\appendices
\section{Proof of proposition~\ref{prop.2}}
\begin{proof}\label{pf.prop2}
As the private data $S$ is independent of noise $R$, we have $\sigma^{2}_{Z}=\sigma^{2}_{S}+\sigma^{2}_{R}$. Let $\gamma=1 /\sigma_{Z}$ and define $Z^{\prime}=\gamma Z$.  Since mutual information is invariant of scaling, we have 
\begin{align*}
    \lim_{\sigma^{2}_{R}\rightarrow \infty}I(S;Z)=\lim_{\gamma \rightarrow 0}I(\gamma S;Z')=0.
\end{align*}
We thus conclude that as long as $\epsilon>0$, there exists noise $R$ with variance $\sigma^{2}_{R}< \infty$ that satisfies $I(S;Z)=\epsilon$.

If the noise $R$ is Gaussian distributed, we can achieve arbitrary small information leakage $I(S;Z)=\epsilon$ as long as $\sigma^{2}_{R}\geq \frac{\sigma^{2}_{S}}{2^{2\epsilon}-1}$. The proof goes as follows:  
\begin{align*}
   \epsilon= I(S;Z)&=h(Z)-h(Z|S)\\
          &=h(Z)-h(R)\\
          &\stackrel{{\text{(a)}}}{=} h(Z)-\frac{1}{2}\log(2\pi e \sigma^2_R) \nonumber\\
          &\stackrel{{\text{(b)}}}{\leq} \frac{1}{2}\log(2\pi e \nonumber \sigma^2_{Z})-\frac{1}{2}\log(2\pi e \sigma^2_R) \nonumber\\
  &=\frac{1}{2}\log(1 +\sigma_{S}^2/\sigma_{R}^2),
\end{align*}
where $h(\cdot)$ denotes the differential entropy; (a) holds as the differential entropy of a Gaussian random variable with variance $\sigma^2$ is given by $\frac{1}{2}\log(2\pi e \sigma^2)$; (b) holds, because the maximum entropy of a random variable with fixed variance is achieved by a Gaussian distribution. 
\end{proof}
\section{Proof of Proposition~\ref{prop.3}}
\begin{proof} \label{pf.prop3}
\begin{align*}
    &I(S_i;\mathcal{V}_i)\\
    &\stackrel{{\text{(a)}}}{=}I(S_i;\{S_j\}_{j\in \mathcal{N}_c}, \tsum_{j \in \mathcal{N}}S_j,\{R_i^j\}_{(i,j) \in\mathcal{E}_c},X^{(0)})\\
    &\stackrel{{\text{(b)}}}{=}I(S_i;\{S_j\}_{j\in \mathcal{N}_c},\{R_i^j\}_{(i,j) \in\mathcal{E}_c},X^{(0)})\\
    &\stackrel{{\text{(c)}}}{=}I(S_i;\{S_j\}_{j\in \mathcal{N}_c},\{R_i^j\}_{(i,j) \in\mathcal{E}_c},\{X_j^{(0)}\}_{j\in \mathcal{N}_h})\\
    &\stackrel{{\text{(d)}}}{=}I(S_i;\{R_i^j\}_{(i,j) \in\mathcal{E}_c},\{X_j^{(0)}\}_{j\in \mathcal{N}_h})\\
    &\stackrel{{\text{(e)}}}{=}I(S_i;\{R_i^j\}_{(i,j) \in\mathcal{E}_c},\{S_j+\tsum_{k\in \mathcal{N}_{j, h}}R_{j|k}+\tsum_{k\in \mathcal{N}_{j, c}}R_{j|k}\}_{j\in \mathcal{N}_h})\\
    &\stackrel{{\text{(f)}}}{=}I(S_i;\{R_i^j\}_{(i,j) \in\mathcal{E}_c},\{S_j+\tsum_{k\in \mathcal{N}_{j, h}}R_{j|k}\}_{j\in \mathcal{N}_h})\\
    &\stackrel{{\text{(g)}}}{=}I(S_i;\{S_j+\tsum_{k\in \mathcal{N}_{j, h}}R_{j|k}\}_{j\in \mathcal{N}_h})\\
    &\stackrel{{\text{(h)}}}{=}I(S_i;\{S_j+\tsum_{k\in \mathcal{N}_{j, h}}R_{j|k}\}_{j\in \mathcal{N}_h'}),
\end{align*}
where (a) holds, as $\forall t\geq 1: S_i\rightarrow X^{(0)}\rightarrow X^{(t)}$ forms a Markov chain (similarly to Remark \ref{rm.1}); (b) holds, as $\sum_{j \in \mathcal{N}}S_j$ can be determined by $X^{(0)}$, i.e., $\sum_{j \in \mathcal{N}}S_j=\sum_{j \in \mathcal{N}}X_j^{(0)}$; (c) holds, as $\{X_j^{(0)}\}_{j\in \mathcal{N}_c}$ can be determined by $\{S_j\}_{j\in \mathcal{N}_c}, \{R_i^j\}_{(i,j) \in\mathcal{E}_c}$ using \eqref{eq.x0SS}, \eqref{eq.rijSS} and \eqref{eq.riSS}; (d) holds because $\{S_j\}_{j\in \mathcal{N}_c}$ is independent of $\{R_i^j\}_{(i,j) \in\mathcal{E}_c},\{X_j^{(0)}\}_{j\in \mathcal{N}_h}$ and $S_i$; (e) holds by representing $\{X_j^{(0)}\}_{j\in \mathcal{N}_h}$ by using \eqref{eq.x0SS} and \eqref{eq.riSS}; (f) follows, as $\{\tsum_{k\in \mathcal{N}_{j, c}}R_{j|k}\}_{j\in \mathcal{N}_h}$ can be determined by $\{R_i^j\}_{(i,j) \in\mathcal{E}_c}$ by using \eqref{eq.rijSS}; 
 (g) holds, as $\{R_i^j\}_{(i,j) \in\mathcal{E}_c}$ is independent of both $S_i$ and $\{S_j+\tsum_{k\in \mathcal{N}_{j, h}}R_{j|k}\}_{j\in \mathcal{N}_h}$; and (h) holds, as $\{S_j+\sum_{k\in \mathcal{N}_{j, h}}R_{j|k}\}_{j\in \mathcal{N}_h\setminus \mathcal{N}_h'}$ is independent of both $S_i$ and $\{S_j+\sum_{k\in \mathcal{N}_{j, h}}R_{j|k}\}_{j\in \mathcal{N}_h'}$.
\end{proof}

\section{Proof of equation \eqref{eq.dospind}}
\begin{proof}\label{pf.dospind}
First consider two successive $\bx$-updates in \eqref{eq.xUp} and plug in \eqref{eq.pdmmLamUp}:
\begin{align}\label{eq.x12}
    \bx^{(t+1)}-\bx^{(t-1)}\nonumber&= \left(\bm I + c\bC^{\top}\bC\right)^{-1}\\
    & \left(- 2c\bC^{\top}\bP\bC\bx^{(t)} - 2c\bC^{\top}\bC\bx^{(t-1)}\right).
\end{align}
We have
\begin{align*}
&I(S_i;\mathcal{V}_i)\\
&\stackrel{{\text{(a)}}}{=}I(S_i;\{S_j\}_{j\in \mathcal{N}_c},\{\Lambda_{i|j}^{(0)}\}_{(i,j) \in\mathcal{E}_c},\{X^{(t)}\}_{t\in\mathcal{T}})\\
&\stackrel{{\text{(b)}}}{=}I(S_i;\{S_j\}_{j\in \mathcal{N}_c},\{\Lambda_{i|j}^{(0)}\}_{(i,j) \in\mathcal{E}_c},\{X^{(t)}\}_{t=1,2})\\
&\stackrel{{\text{(c)}}}{=}I(S_i;\{S_j\}_{j\in \mathcal{N}_c},\{\Lambda_{i|j}^{(0)}\}_{(i,j) \in\mathcal{E}_c},\{X_j^{(t)}\}_{j\in \mathcal{N}_h,t=1,2})\\
&\stackrel{{\text{(d)}}}{=}I(S_i;\{S_j\}_{j\in \mathcal{N}_c},\{\Lambda_{i|j}^{(0)}\}_{(i,j) \in\mathcal{E}_c}\\
& \hspace{1cm},\{S_j-{\tsum_{k\in \mathcal{N}_{j, h}}} \bB_{j|k}\Lambda^{(t)}_{k|j}\}_{j\in \mathcal{N}_h,t=0,1})\\
&\stackrel{{\text{(e)}}}{=}I(S_i;\{S_j-{\tsum_{k\in \mathcal{N}_{j, h}}} \bB_{j|k}\Lambda^{(t)}_{k|j}\}_{j\in \mathcal{N}_h,t=0,1}
\\
& \hspace{1cm}|\{S_j\}_{j\in \mathcal{N}_c},\{\Lambda_{i|j}^{(0)}\}_{(i,j) \in\mathcal{E}_c})
\end{align*}
where (a) holds, as all $\{\Lambda_{i|j}^{(t>0)}\}_{(i,j) \in\mathcal{E}_c}$ can be determined by $\{X^{(t)}\}_{t\in\mathcal{T}}$ and $\{\Lambda_{i|j}^{(0)}\}_{(i,j) \in\mathcal{E}_c}$ from \eqref{eq.lamdaUp}; (b) holds, as all $\{X^{(t)}\}_{t>2}$ can be determined by $\{X^{(t)}\}_{t=1,2}$ on the basis of \eqref{eq.x12} (note that we omit $X^{(0)}$ by assuming $\bx$ is initialized with all zeros); (c) holds, as $\{X_j^{(2)}\}_{j\in \mathcal{N}_c}$ can be constructed by using $\{S_j\}_{j\in \mathcal{N}_c}, X^{(1)}, \{\Lambda_{i|j}^{(1)}\}_{(i,j) \in\mathcal{E}_c}$ based on \eqref{eq.xiup}, in which the last set can be determined on the basis of (a), and similarly $\{X_j^{(1)}\}_{j\in \mathcal{N}_c}$ can be constructed by $\{S_j\}_{j\in \mathcal{N}_c},\{\Lambda_{i|j}^{(0)}\}_{(i,j) \in\mathcal{E}_c}$; (d) also follows from \eqref{eq.xiup}; and (e) follows from the definition of conditional mutual information and $S_i$ being independent of both $\{S_j\}_{j\in \mathcal{N}_c}$ and $\{\Lambda_{i|j}^{(0)}\}_{(i,j) \in\mathcal{E}_c}$.
\end{proof}
\bibliographystyle{IEEEbib}
\bibliography{dualpath}

\end{document}